\documentclass[%
 reprint,
 amsmath,amssymb,
 aip,
]{revtex4-2}

\usepackage{graphicx}% Include figure files
\usepackage{dcolumn}% Align table columns on decimal point
\usepackage{bm}% bold math
\usepackage[]{color}
\usepackage{subfig}
\begin{document}

\title {Influence of water models on water movement through AQP1}
% Force line breaks with \\

% \author{Miguel A. Gonzalez}
% \email{miguelangel.gonzalez@ucm.es}
%  \altaffiliation[Now at ]{Departamento de Qu\'imica F\'isica, Universidad Complutense de Madrid, 28040, Madrid ES}%Lines break automatically or can be forced with \\
% \author{Fernando Bresme}%
%  \email{f.bresme@imperial.ac.uk}
% \affiliation{%
% Department of Chemistry, Imperial College London, SW7 2AZ, London UK
%  }%

% \collaboration{MUSO Collaboration}%\noaffiliation

% \affiliation{
% \affiliation{
% \affiliation{%
%  Authors' institution and/or address\\
%  This line break forced with \textbackslash\textbackslash
% }%


% \date{}% It is always \today, today,

\author{Miguel A. Gonzalez}
\email{miguelangel.gonzalez@urjc.es}
\affiliation{Universidad Rey Juan Carlos, ESCET, 28933, Mostoles-Madrid, Spain}

\author{Alberto Zaragoza}
\affiliation{Department of Chemistry, The University of Utah, Salt Lake City, Utah 84112-0850, USA}
\author{Charlotte I. Lynch}
\affiliation{University of Oxford, Department of Biochemistry, South Parks Road, OX1 3QU, Oxford, UK}
\author{Mark S.P. Sansom}
\affiliation{University of Oxford, Department of Biochemistry, South Parks Road, OX1 3QU, Oxford, UK}
\author{Chantal Valeriani}
\affiliation{Universidad Complutense de Madrid, Facultad de Ciencias Fícias, Departamento de Estructura de la Materia, Física Térmica y Electrónica, 28040, Madrid, Spain}

\begin{abstract}
  Water diffusion through membrane proteins is a key aspect of cellular function. Essential processes of cellular metabolism are driven by osmotic pressure, which depends on water channels. Membrane proteins such as aquaporins (AQPs) are responsible for enabling water permeation through the cell membrane. AQPs are highly selective, allowing only water and relatively small polar molecules to cross the membrane. 
  Experimentally, estimation of  water flux through membrane proteins is still a challenge, and hence accurate simulations of water permeation are of particular importance. We present a numerical study of water diffusion through AQP1 comparing three water models: TIP3P, OPC and TIP4P/2005. Bulk diffusion, diffusion permeability and osmotic permeability are computed and compared among all models. The results show that there are significant differences between TIP3P (a particularly widespread model for simulations of biological systems), and the more recently developed TIP4P/2005 and OPC models. We demonstrate that  OPC and TIP4P/2005 reproduce protein-water interactions and dynamics in very good agreement with experimental data. From this study, we find that the choice of the water model has a significant effect on the computed water dynamics as well as its molecular behaviour  within a biological nanopore.
\end{abstract}
\maketitle

\section{Introduction}
Water is the most abundant molecule  in cells.
Even though water is able to diffuse slowly through a lipid bilayer, for essential physiological processes a high water flux is required.
This can be achieved making use of protein membranes such as aquaporins
(AQPs) which allow for a high selectivity for water permeation across biological membranes~\cite{Beitz2009,Yang2017}.
AQPs are located in several cells found in  the brain, kidneys, skin, blood vessels, liver and connective tissue.~\cite{Castle2005,Day2014}
The lack of functionality of such cells might induce several diseases, such as glaucoma, brain edema or congestive heart failure.~\cite{Castle2005,Kruse2006,Agre2003} 
Although evidence of the existence of water channels had been reported several years earlier, the aquaporin structure was not described until 1992~\cite{Preston1992}.
All members of the AQP family are small ($\sim$ 30~kDa) intrinsic membrane proteins with a strongly hydrophobic character, and consist of four monomers. 
Each monomer is a functional water channel having a hourglass shape~\cite{Walz1997} with an amphiphilic interior and a  selectivity filter.
The primary sequence of those proteins shows six trans-membrane helices (I-VI) linked by five loops.
The water flux inside the channel depends on an asparagine-proline-alanine (NPA) motif,  found in two loops located approximately in the middle of the membrane protein channel.
The interaction between these motifs not only confers a particular shape to the channel but it is also responsible for the orientation of water molecules inside the channel. 
In combination with the NPA motif, water permeation is also modulated by the aromatic/arginine selectivity filter which is situated next to the extracellular exit/entrance of the channel.~\cite{Chen2006,Eriksson2013}   
The topology of the pore is extremely important as it enables selected small polar molecules to cross the membrane (e.g. H$_{2}$O, NH$_{3}$, glycerol) whilst preventing the passage of others (e.g. ions and larger molecules). 
Based on its permeability, AQPs can be classified in two subfamilies: classical aquaporins (permeable to water) and aquaglyceroporins (permeable to both water and glycerol).

AQP1 is a classical aquaporin with a high water permeability
as described in detail in Ref.~\cite{Tajkhorshid2002} and confirmed in several numerical studies~\cite{DeGroot2001,Chen2006,Eriksson2013,Law2014,Ariz-Extreme2017,Decker2017}.  
There are three different approaches presented in the literature to study water through aquaporins: 1) a continuum hydrodynamic approach~\cite{Gravelle2013} that performs surprisingly well for a nanopore of molecular dimensions, capturing key steric effects; 2) a potential of mean force approach, to study the energetics of water transport~\cite{Aponte-Santamaria2017,Ariz-Extreme2017}; 3) ``a continuous-time random-walk model" approach of water transport across the channel~\cite{Berezhkovskii2002}.

As suggested by De Groot \textit{et al.} in Ref.~\cite{Grubmulle2005}, approaches 2 and 3) are in a quantitative agreement. Thus, we will follow approach 3 throughout our study (extracting equations from Ref.~\cite{Zhu2004}).
Numerous molecular dynamics studies have been conducted on the water permeation mechanism through aquaporins~\cite{DeGroot2001,DeGroot2001a,DeGroot2003,Hashido2005,Hub2006,Hub2010,Mamonov2007,Jensen2003,Tajkhorshid2002,Zhu2002,Lee2016}.
As  discussed by Ozu \textit{et al.}~\cite{Ozu2013}, many AQP simulations have been carried out using different force fields (GROMOS or CHARMM) for the protein and the standard older TIP3P or SPC water models.
De Groot and Grubm\"uller~\cite{DeGroot2001a} demonstrated that proton translocation across AQPs was prevented by an electrostatic barrier.
However Jensen \textit{et al.}~\cite{Jensen2003}, using a different force field to simulate AQP and water (TIP3P), concluded that the proton induced an interruption of the hydrogen bond chain network built into the channel, preventing the proton from crossing the pore. 
Thus, it is reasonable to suggest that models of water permeation might be affected by the force field used to simulate water molecules. 

When dealing with pure water, both Vega \textit{et al.}~\cite{vega11} and Onufriev \textit{et al.}~\cite{Onufriev2018} carried out a detailed comparison between experimental and numerical values of  thermodynamic and dynamic properties, simulating water by means of several interaction potentials. 
Even though they clearly demonstrated that TIP3P was not one of the best performing potentials for pure water, TIP3P  remains a widely used water model in biomolecular simulations, in combination with CHARMM (the most widely used protein force field).
With this in mind, we propose to study the molecular mechanism of water permeation of AQP1 (1H6I code PDB)~\cite{DeGroot2001} simulated with CHARMM36m~\cite{Huang2017}, in combination with the TIP3P, OPC and TIP4P/2005 water models. Even though the latter two water models are less frequently used in biomolecular simulations, they are  known to give very reliable thermodynamic and dynamic properties for bulk water. 
To our knowledge, TIP4P/2005 has been already implemented with the AMBER ff03w~\cite{Best2010} force field as a default water model but
has not been employed in simulations of AQP,  where the microscopic behaviour of water clearly plays a vital role.
It is important to note that we have chosen to focus our study on rigid non-polarizable water models. There is a range of potentially more accurate polarizable and flexible water models, however the use of these comes at significant computational cost~\cite{Charlotte2020}. It is therefore beneficial to first assess the consistency and accuracy of a range of rigid non-polarizable models in order to establish whether the use of polarizable water models is needed for the study of water behaviour in aquaporins. Similar comparisons of non-polarizable water models have been conducted for the 5-HT3 receptor channel~\cite{Klesse2020} and Cucurbit[7]uril-guest systems~\cite{cinaroglu2021}.
In this article, we have studied several properties of human AQP1 (water diffusion across the pore, diffusion permeability and osmotic permeability) taking into account the molecular mechanism characteristic of AQP.
In this way we have selected to simulate a range of representative (non-polarizable) models of water, comparing two more recent water potentials to what is still the ``standard model". This allows us to capture the likely range of behaviours for this class of model.

\section{Methods}

\subsection{Simulation details}

\begin{figure}
    \centering
a)\includegraphics[width=0.4\linewidth]{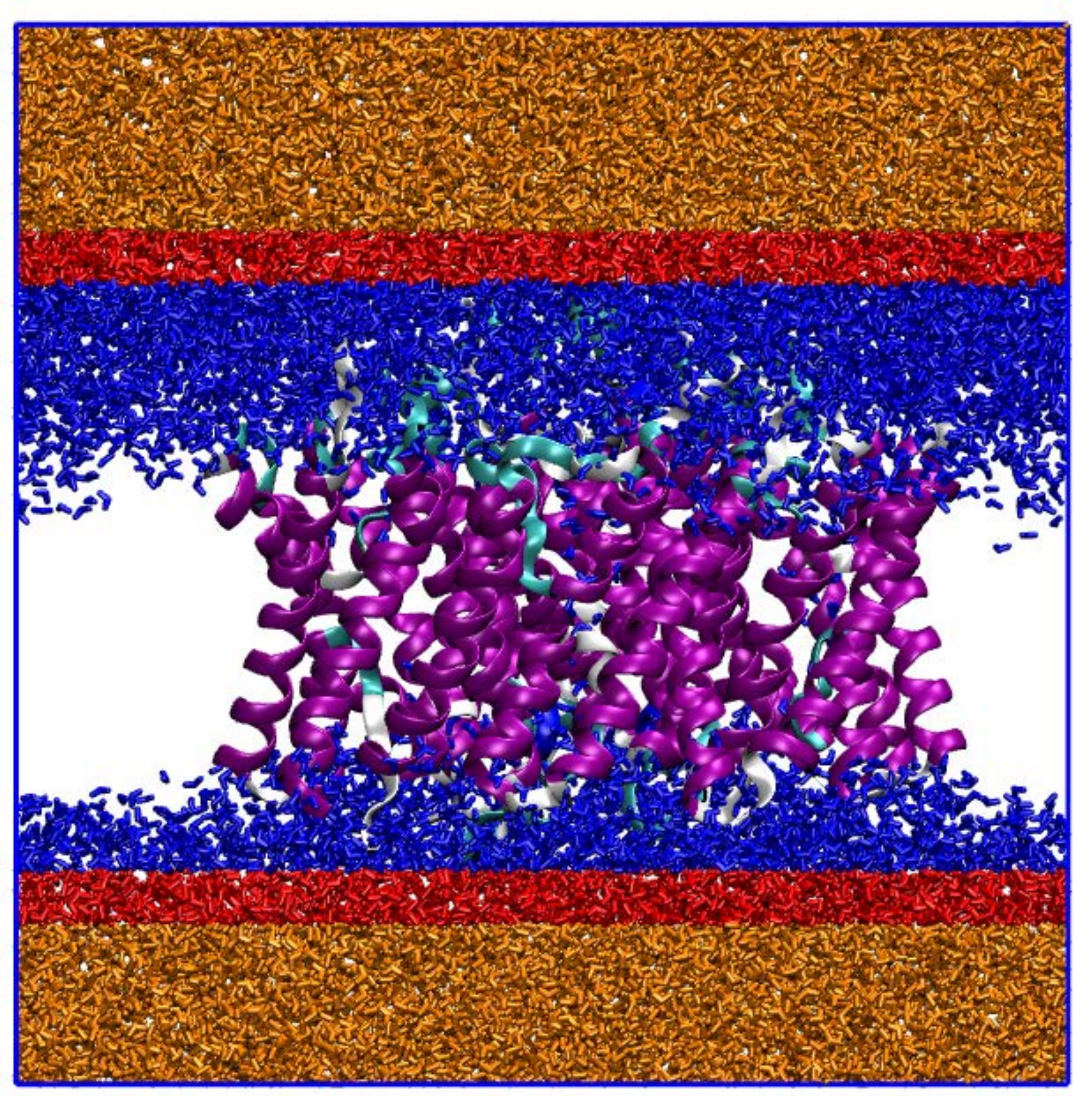}
b)\includegraphics[width=0.4\linewidth]{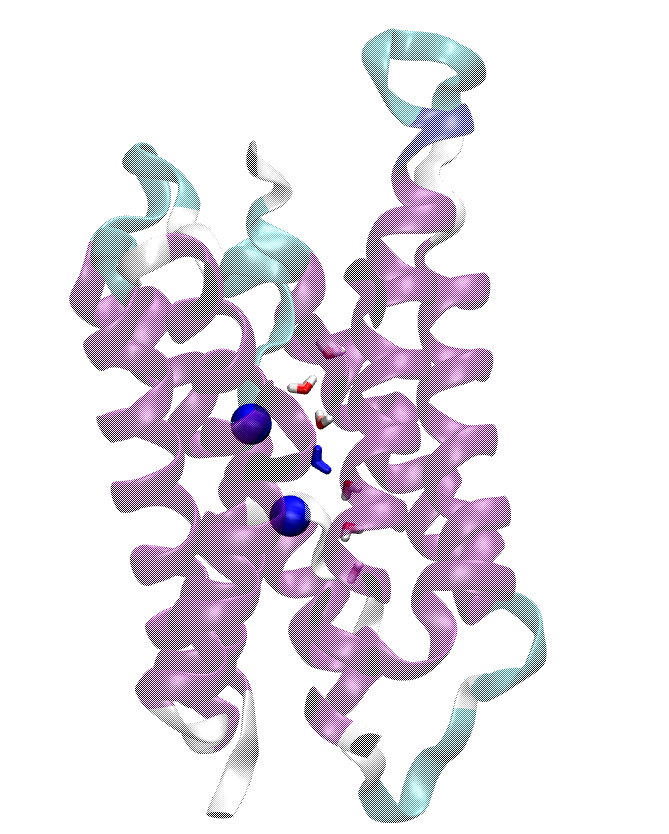}
    \caption{a) Rendered image of the simulated system (from top/bottom towards the middle): bulk water molecules (in brown), interfacial water molecules (in red), lipids (in blue) and      AQP1 (in purple). b) Rendered image of the AQP1 channel, showing water molecules passing through, while changing their dipolar moment when crossing the center. The water molecules are plotted in red/white.}
    \label{fig:snapAQP1}
\end{figure}

We have carried out molecular dynamics simulations for  AQP1 embedded in a palmitoyloleoylphosphatidylcholine (POPC) bilayer membrane using the GROMACS 2016.4 software package~\cite{Abraham2015}.
Human AQP1 (1H6I code PDB)~\cite{DeGroot2001} is simulated via the CHARMM36m force field~\cite{Huang2017}, whilst the water molecules are simulated using TIP3P~\cite{jorgensen81}, OPC~\cite{Izadi2014}, and TIP4P/2005~\cite{abascal05b} potentials.
In order to maintain the protein fold, the simulations were carried out with backbone restraints.
Figure \ref{fig:snapAQP1} a) shows an  image of the system under study, representing the different regions where water molecules can be found: bulk water molecules (in brown), interfacial water molecules (in red), lipids (in blue) and AQP1 (in purple). Panel b) represents a rendered image of the AQP1 channel, showing water molecules (in red/white) passing through, while changing their orientation when crossing the center.

Considering the three models of water, we have set up three initial configurations with the AQP1 tetramer embedded in a lipid bilayer of 209 POPC molecules and  solvated with approx.~20000 water molecules.
In order to prepare a neutral configuration, we have added 4 Cl$^{-}$ ions.
Each AQP1-water system was simulated for 100~ns, and simulations were repeated four times  with different random initial velocities. 
We set a time step of 1~fs keeping the temperature constant at 298~K using a velocity rescale thermostat~\cite{bussi07} with a relaxation time set to 0.67~ps. 
The Parrinello-Rahman barostat~\cite{parrinello81} was applied to maintain the pressure constant at 1~bar using a 2.57~ps  relaxation time. 
When dealing with water, the Lennard-Jones (LJ) potential was truncated at 1.2~nm, adding standard long-range corrections to the LJ energy.
Ewald sums~\cite{essmann95} were considered to compute long-range electrostatic forces with a real space cut-off at 1.2~nm.
In order to compare the behaviour of the three water models and their interactions with AQP1, we have selected the most relevant properties that allow us  to establish the influence of water potentials on water dynamics through protein channels.

\subsection{Diffusion, diffusion permeability, rate of water molecules and average dipole moment across the channel}
Water diffusion through a cell membrane is considered as a subdiffusive process~\cite{Rog2002,Bhide2005,Yamamoto2013,Yang2014,Toppozini2015}.
Also, Yamamoto \textit{et al.}\cite{Yamamoto2015} demonstrated that the water diffusion near a lipid membrane is lower than the bulk due to two mechanisms. On the one hand, there is a divergent mean trapping time induced by the lipid-water interaction which can be described by a continuous-time random walk. On the other hand, the viscoelasticity of the lipid membrane induces a long-time correlated noise described by fractional Brownian motion.
In order to compute  water diffusion, we distinguished four regions depending on the distance ($d$) between the lipid headgroup and the position of the water molecule. 
As in Ref.~\citenum{Yang2014}, the criteria applied to create those regions are the following: we identify a 1) bulk region whenever $d$ $>$ 0.6~nm, an 2) interface region whenever 0.3~nm $< d <$ 0.6~nm, a 3) contact region whenever $d <$ 0.3~nm) and an 4) AQP region defined as the region inside the lipid headgroups (corresponding to waters in the pore).
The long time diffusion coefficient ($D$) is then  computed  using Einstein's relation based on the water mean square displacement: 
%%%%%%%%%%%%%%%%%%%%%%%%%%%%%%%%%%%%%%%%%%%%%%
\begin{equation}
\label{eq:diffu_constant}
    D=\frac{\left \langle \left | r(t_{0}+t)-r(t_{0}) \right |^{2} \right \rangle}{2\cdot dim \cdot  t}
\end{equation}
%%%%%%%%%%%%%%%%%%%%%%%%%%%%%%%%%%%%%%%%%%%%%%
where $r(t)$ is the position of the water molecule centre of mass, $t$ is the time, and $dim$ the dimension that depends on the geometry of the system that the molecules are moving through, with $dim=1$ for 1D geometries such as carbon nanotubes CNT or AQP1 and $dim=3$ for bulk~\cite{Zaragoza2019}.   

Another approach to calculate water diffusion inside the protein channel (AQP region)  is based on a single-water molecule column~\cite{Berezhkovskii2002}.
In this case, water diffusion ($D_{z}$) is computed according to the  relation~\cite{Horner2015}:
%%%%%%%%%%%%%%%%%%%%%%%%%%%%%%%%%%%%%%%%%%%%%%
\begin{equation}
\label{eq:diffu_z}
    D_{z}=\frac{k_{0}z^{2}}{2}=\frac{z^{2}p_{f}}{2\nu_{w}}
\end{equation}
%%%%%%%%%%%%%%%%%%%%%%%%%%%%%%%%%%%%%%%%%%%%%%
where $k_{0}$, $z$, $p_{f}$ and $\nu_{w}$ are the rate of water molecules across the channel, the average distance between two water molecules  inside the channel, the osmotic permeability (see below for more details) and the volume of a single molecule, respectively. 

Water permeability can be divided into osmotic permeability ($p_{f}$) and diffusion permeability ($p_{d}$) depending on the system's conditions. 
Osmotic pressure drives the movement of water molecules through the membrane protein when there is a solute gradient across the membrane.
Hence in these conditions water permeability would be given by the osmotic permeability of the system.

As employed in previous studies,  
\cite{Shahbabaeia2020,Grubmulle2005,Zhu2004}
osmotic permeability is defined as the proportionality constant between the net water flux ($j_{w}$) across the protein channel and the solute concentration difference ($\Delta C_{s}$), 
%%%%%%%%%%%%%%%%%%%%%%%%%%%%%%%%%%%%%%%%%%%%%%
\begin{equation}
\label{eq:osmopermea1}
    j_{w}=p_{f}\Delta C_{s}.
\end{equation}
%%%%%%%%%%%%%%%%%%%%%%%%%%%%%%%%%%%%%%%%%%%%%%
Taking into account the rate of water molecules across the membrane protein, $k_{0}$, the osmotic permeability is also defined as 
%%%%%%%%%%%%%%%%%%%%%%%%%%%%%%%%%%%%%%%%%%%%%%
\begin{equation}
\label{eq:osmopermea2}
    p_{f}=\nu_{w}k_{0}.
\end{equation}
%%%%%%%%%%%%%%%%%%%%%%%%%%%%%%%%%%%%%%%%%%%%%%
At equilibrium and in the absence of solute or solute gradient, there is still net water moving through the channel that could be attributed to diffusion of water molecules due to temperature.
Thus, the diffusion permeability, $p_{d}$, allows us to study the water flux across the protein under these conditions: 
%%%%%%%%%%%%%%%%%%%%%%%%%%%%%%%%%%%%%%%%%%%%%%
\begin{equation}
\label{eq:diffpermea1}
    j_{tr}=p_{d}\Delta C_{tr}
\end{equation}
%%%%%%%%%%%%%%%%%%%%%%%%%%%%%%%%%%%%%%%%%%%%%%
where $\Delta C_{tr}$ corresponds to the difference in the tracer's concentration  between the two sides of the membrane and $j_{tr}$ is the net tracer's flux (where a tracer is defined as a labelled water molecule that can be distinguished from similar molecules, \textit{i.e.} a molecule of heavy water is a tracer for water).
At equilibrium, the rate of water molecules crossing the channel is computed by $q_{0}$, which is  related to $p_{d}$ via  
%%%%%%%%%%%%%%%%%%%%%%%%%%%%%%%%%%%%%%%%%%%%%%
\begin{equation}
\label{eq:diffpermea2}
    p_{d}=(V_{w}/N_{a})q_{0}=\nu_{w}q_{0}.
\end{equation}
%%%%%%%%%%%%%%%%%%%%%%%%%%%%%%%%%%%%%%%%%%%%%%
Note that the centre used to decide if a water molecule crossed the channel is the average of Z position of N atoms for ASN residue inside the  channel. 
One molecule has been considered that it crossed the AQP when its Z coordinate is 1.5 nm from this point. This centre is updated each single time step during the analysis.

Molecular simulations become an essential tool for computing $p_{d}$ as they allow us to label individual water molecules and therefore distinguish them from one another.
Both properties, $p_{d}$ and $p_{f}$, are proportional to the number of water molecules, $N$, taking part in the translocation following the continuous-time random-walk process~\cite{Berezhkovskii2002} based on Ref.~\cite{Finkelstein1987}.
For AQP1, Zhu \textit{et al.}~\cite{Zhu2004} presented a value for $p_{f}/p_{d}=N+1=11.9$. 
This is in good agreement to the experimental value of $p_{f}/p_{d}=13.2$~\cite{Mathai1996}.
However, these values are higher than the average number of molecules inside the protein channel, since there are molecules of water outside the channel that participate in the permeation across the membrane.

In order to compute the flux in Eq.~\ref{eq:diffpermea1} we need to compute $p_d$ and $\Delta C_{tr}$. 
The latter is computed by simple counting whereas the former is computed via Eq.~\ref{eq:diffpermea2}.  
To compute $q_{0}$ in Eq.~\ref{eq:diffpermea2} we count the number of water molecules that cross the membrane through the AQP1. Thus, we define the centre of the protein and a typical distance from which a molecule is considered to be outside  the channel.
The  channel centre is identified in the following way: two nitrogen atoms inside the channel are bound to water via hydrogen bonds and their orientation is essential to describe the AQP1 permeability.
Thus, we compute the channel centre as the average of these nitrogen atom positions. 
The channel length is set at 1.5~nm from the centre.

Water molecules in bulk are characterised by an average dipole moment equal to zero. However, in nano-confined structures such as a CNT~\cite{Zaragoza2019} or membrane channels~\cite{Tajkhorshid2002}, there is a particular orientation of the molecules due to the hydrogen bond network.   
Thus, the dipole moment is an important parameter to study the molecular orientations inside the channel, both artificial and natural. 
In our work, we have computed the water dipole moment $\mu$  according to   $\vec{\mu}=\sum_{i}q_{i}\vec{r_{i}}$, where  $q_{i}$ and $r_{i}$ are the charge and the position vector for molecule $i$, respectively.
We focus on the $z$ component of $\vec{\mu}$, since in our simulations the orientation of the water molecule inside the channels is mainly projected onto this axis. 
Hence, we calculate the average dipole moment of slabs, having a thickness of 0.025~nm, across the $z$ axis to compare the molecular orientation between the channel and molecules.

\section{Results And Discussion}

\subsection{Water molecule orientations inside  AQP1}

As suggested in Ref.\cite{Tajkhorshid2002}, water molecules arrange inside the channel forming a single-molecule chain with a specific orientation. 
In our work we demonstrate that this mechanism is reproduced by the three water models. 

\begin{figure}[h!]
\centering
\includegraphics[width=0.6\linewidth]{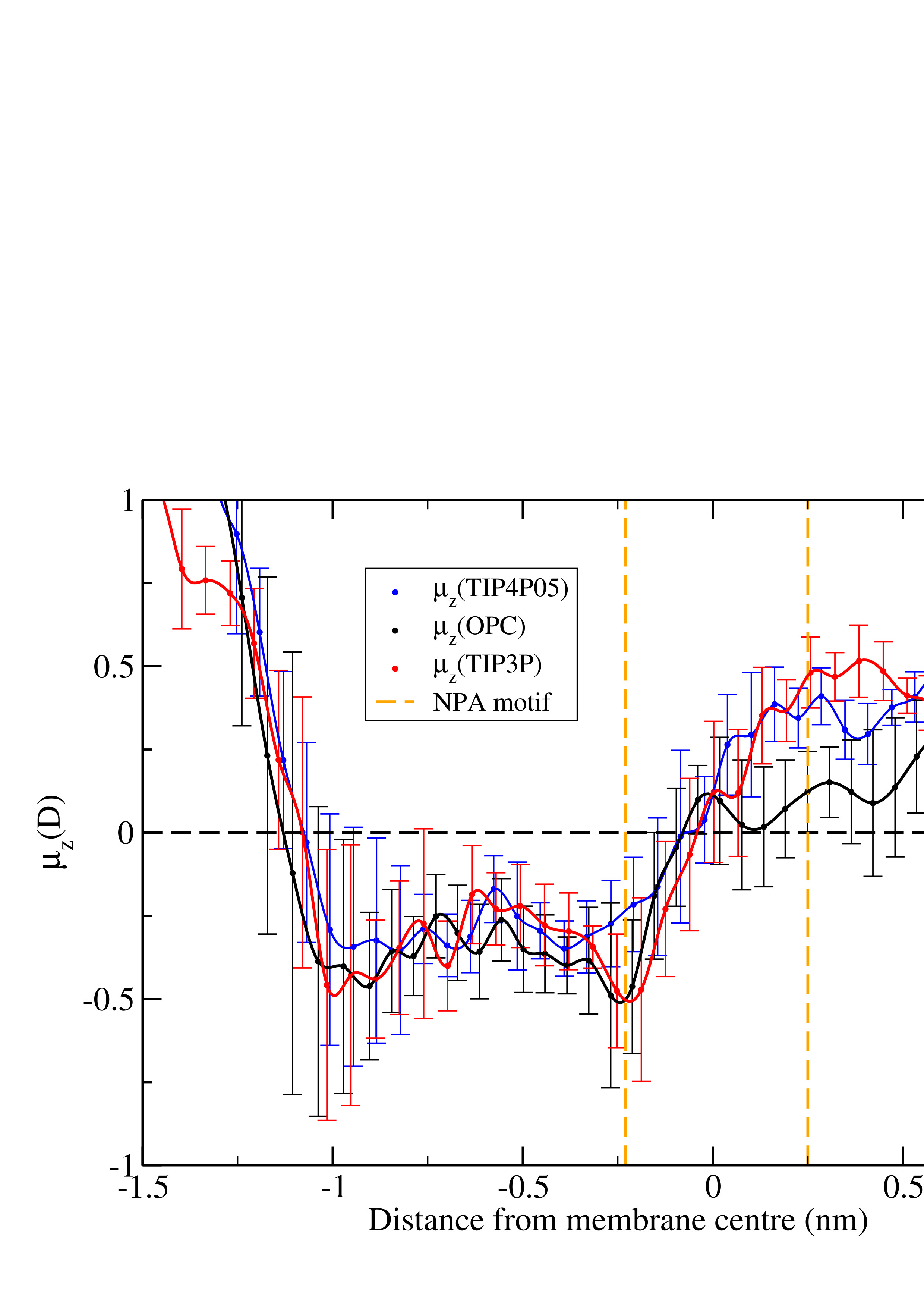}
\caption{ $z$ component of the average dipole moment $\mu_{z}$  for the three water models. The red line is for TIP3P, the blue line  for TIP4P/2005, and the black line  for OPC. Error bars have been added taking into account the standard deviation of $\mu_{z}$. The vertical dashed lines indicate the approximate locations of the
Asn residue of the NPA motifs.
Note that  the centre of the bilayer is at z = 0 nm, and the pore runs from approximately z = -1 to +0.8~nm.}
\label{fig:dipole_moment}
\end{figure}

As shown in  Fig.~\ref{fig:dipole_moment}, the water molecule orientation can be illustrated by computing the average of the $z$ component of the dipole moment  inside the channel. 
%%%%%%%%%%%%%%%%%%%%%%%%%%%%%%%%%%%%%%%%%%%%%%%%%%%%%%%%%%%%%
%%%%%%%%%%%%%%%%%%%%%%%%%%%%%%%%%%%%%%%%%%%%%%%%%%%%%%%%%%%%%
Although there is a dynamic permeation across the channel, it is possible to distinguish the positions of water molecules, described by the slightly ragged shape of the water dipole moment inside the protein (see Fig.~\ref{fig:dipole_moment} ).   
These results demonstrate that there are two stable orientations inside the channel as described by Ref.~\cite{Tajkhorshid2002}.

There are no major differences between the dipole moment profiles for the three water models, although the degree of water dipole orientation is less for the OPC model in the second (i.e. $z >0$) half of the pore than for the other two water models.
This implies that, even though TIP3P is the only water model fully ``tuned'' with CHARMM36, both OPC and TIP4P/2005 are capable of capturing a qualitatively similar behaviour inside the AQP. Therefore, one could deduce that the similarity between the dipole moments computed with the different water models is due to water interacting in a similar way with AQP1.
Water interacts with AQP via hydrogen bonds that can be only formed when water is close enough to the asparagines located at the middle of the pore. Our results on the dipolar moment  suggest that water interacts with asparagine in a similar way independently of the model used to mimic its behaviour.

\subsection{Water molecule diffusion through the protein channel} \label{subsec:diffusion}
As in Ref.\citenum{Yang2014}, we identify  interfacial, pore and contact regions, depending on the position of the water molecule with respect to the bilayer and the AQP1 pore. 
To compute the water diffusion constant ($D$)  we use Eq.~\ref{eq:diffu_constant} with $dim=3$ for the bulk ($D_{bulk}$), interface ($D_{inter}$), and contact ($D_{contact}$) regions, and $dim=1$ for the AQP ($D_{AQP}$) region (AQP1  being almost cylindrical). 
%%%%%%%%%%%%%%%%%%%%%%%%%%%%%%%%%%%%%%%%%%%%%%%%%%%%%%%%%%%%%
\begin{figure*}[h!]
\subfloat{
\includegraphics[clip,width=0.3\textwidth]{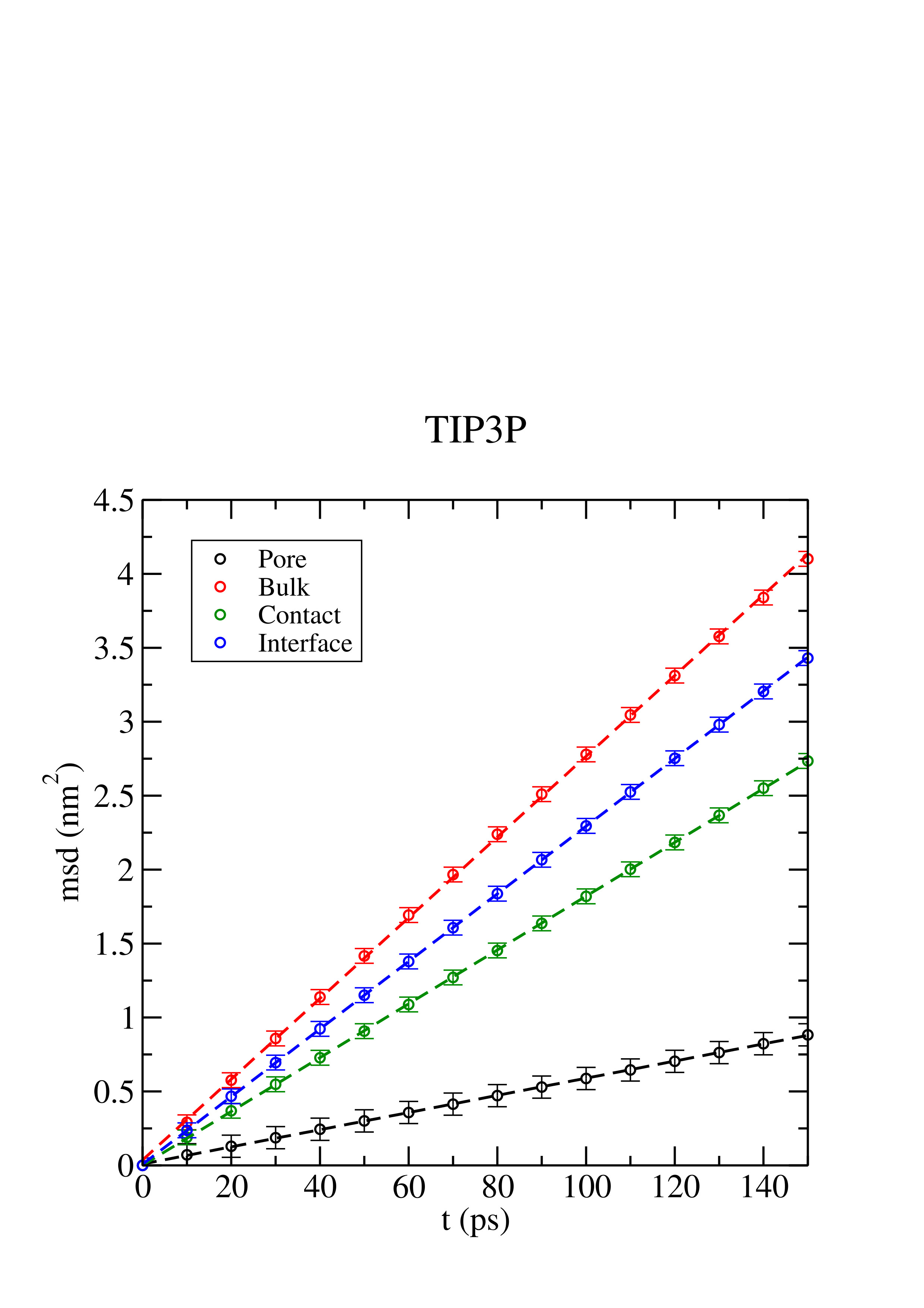}}
\hspace{\fill}
\subfloat{
\includegraphics[clip,width=0.3\textwidth]{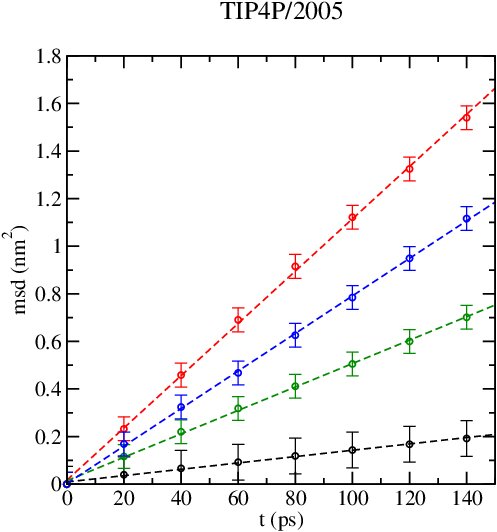}}
\hspace{\fill}
\subfloat{
\includegraphics[clip,width=0.3\textwidth]{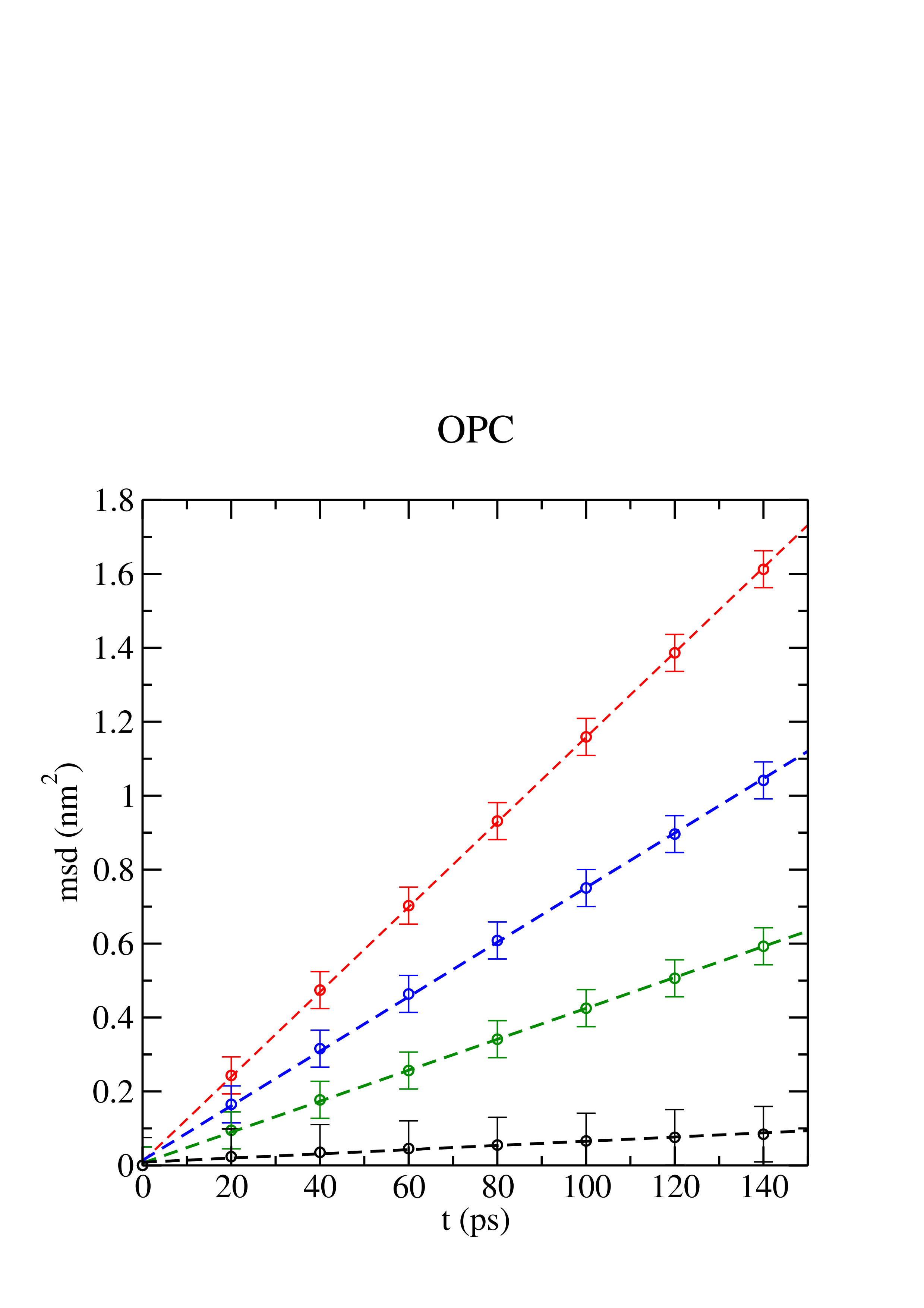}}\\
\caption{ Mean square displacement (nm$^{2}$) for TIP3P (left), TIP4P/2005 (centre), and OPC (right). 
The coloured lines denote the following regions: 
bulk (red), interface  (blue), lipid-water contact region (green), and AQP (black).}
\label{fig:msd}
\end{figure*}
Moreover, we compare the value of the diffusion computed in the bulk region at $d > 0.6$~nm ($D_{bulk}$) to the one computed for each model in bulk ($D_{model}$), since the presence of the lipid membrane reduces the value of $D$ in nearby regions (even at distances higher than 2~nm~\cite{Yamamoto2015}). 
%%%%%%%%%%%%%%%%%%%%%%%%%%%%%%%%%%%%%%%%%%%%%%%%%%%%%%%%%%%%%
All  mean square displacements (msd) for each region are reported  in Fig.~\ref{fig:msd}. 
Two groups of data could be distinguished for each model (although in the case of TIP3P it is more evident). 
The bulk, interface and contact regions have similar slopes, while the slope for the AQP region is lower than the rest. 
The diffusion coefficient (D) values for all water models in bulk and in the various  regions are presented in Table \ref{tab:diffuregions} (as   calculated from the mean square displacement in Fig.~\ref{fig:msd}).

%%%%%%%%%%%%%%%%%%%%%%%%%%%%%%%%%%%%%%%%%%%%%%%%%%%%%%%%%%%%%
\begin{table*}[ht!]
\caption{\label{tab:diffuregions}Diffusion coefficient values computed using Eq.~\ref{eq:diffu_constant} with $dim=3$ in the bulk, interface  and contact region, with $dim=1$ in the AQP region for TIP3P, TIP4P/2005 and OPC water models. The values in parenthesis are the standard deviations. All diffusion coefficients are expressed in 10$^{-5}$ (cm$^{2}$/s).}
\begin{ruledtabular}
%\centering
%\begin{threeparttable}
\begin{tabular}{lcccc}
Water model & D$_{bulk}$ & D$_{inter}$ & D$_{contact}$& D$_{PORE}$\\ 
\hline
TIP3P      & 4.60 (0.01) & 3.74 (0.02) & 2.93 (0.02) & 0.10 (0.01) \\ 
TIP4P/2005 & 1.87 (0.04) & 1.24 (0.1)  & 0.77 (0.04) & 0.05 (0.02) \\ 
OPC        & 1.94 (0.04) & 1.26 (0.08) & 0.72 (0.09) & 0.02 (0.02) \\ 
Exp.       & 2.30~\cite{vega11} & & & 0.04\\
\end{tabular}
\end{ruledtabular}
\end{table*}
%%%%%%%%%%%%%%%%%%%%%%%%%%%%%%%%%%%%%%%%%%%%%%%%%%%%%%%%%%%%%%%%%%%%%

When dealing with bulk water, the literature values of $D_{literature}$ at 298K and 1 bar are $5.5$x$10^{-5}$~cm$^2$/s~\cite{MacKerell1998} for TIP3P, $2.06$x$10^{-5}$~cm$^2$/s for TIP4P/2005~\cite{abascal05b} and  $2.3$x$10^{-5}$~cm$^2$/s~\cite{Izadi2014} for OPC $D_{literature}$.
These data are higher than the $D_{bulk}$ computed from the mean square displacement in our simulations (see Fig.~\ref{fig:msd}), reflecting the phenomena of subdiffusion of water close to the lipid membrane, explained in Ref.~\cite{Yamamoto2015}, and in addition it might be due to differences in the simulation conditions, such as the value of the cut-off and/or the thermostat applied during the simulation. However, the value of water diffusion given in this article is reliable.
However, it is clear that TIP4P/2005 and OPC water models reproduce the experimental bulk water diffusion better than TIP3P.
Concerning the hydrogen bonds formed in bulk water, a recent paper\cite{martelli2021} has shown that 
the TIP3P water model yields a higher rotational relaxation time and a lower network complexity  parameter. This results in a higher diffusion coefficient for TIP3P water in bulk. 
In contrast, TIP4P/2005 (a water model very similar to OPC) yields a lower rotational relaxation time and a higher network complexity parameter. This results in a lower diffusion coefficient for TIP4P/2005 (and by extension for OPC) water in bulk. 
TIP3P bulk water has a higher diffusion coefficient than both OPC and TIP4P/2005, due to its higher rotational relaxation time and lower network complexity  parameter. This effect is enhanced when water is confined within the single file pore AQP, as already suggested in Ref\cite{Zaragoza2019}.

On the one hand, although it is possible to consider TIP3P water  subdiffusing  as $D$ is compared to $D_{literature}$, TIP3P  overestimates  bulk water diffusion in comparison to the experimental data of $2.3$x$10^{-5}$~cm$^2$/s in all regions.
On the other hand, TIP4P/2005 and OPC show a consistent subdiffusion of water molecules  close to the membrane, as expected~\cite{Yamamoto2015} both  with respect to their $D_{literature}$ and in comparison to the experimental value.
Focusing on the AQP region, \textit{i.e.} inside the membrane protein, the experimental data of water diffusion is $0.04$x$10^{-5}$~cm$^2$/s~\cite{Horner2015}.
TIP3P gives $0.10$x$10^{-5}$~cm$^2$/s for $D_{PORE}$, which is dramatically higher than the experimental value and also higher than the other water models,  $0.05$x$10^{-5}$~cm$^2$/s and $0.02$x$10^{-5}$~cm$^2$/s for TIP4P/2005 and OPC, respectively.
When comparing the diffusion coefficient of water inside the pore with respect to its bulk counterpart, we get a comparable ratio independent of the model.                        
The reason is that TIP3P water diffuses more than the other water models not only in bulk but also when confined in a nanopore.
In addition, note that the diffusion of water inside the channel is different with respect to the experimental value depending on the chosen model. In the case of TIP3P the diffusion is 2.5 times higher than the experimental one. OPC also shows a significant discrepancy, although in this case the model presents a diffusion that is 2 times lower than the experimental one. However, the diffusion of the TIP4P/2005 model is in optimal agreement with the experimental value.  

\begin{figure*}[ht!]
\centering
\subfloat{
\includegraphics[clip,width=0.3\textwidth]{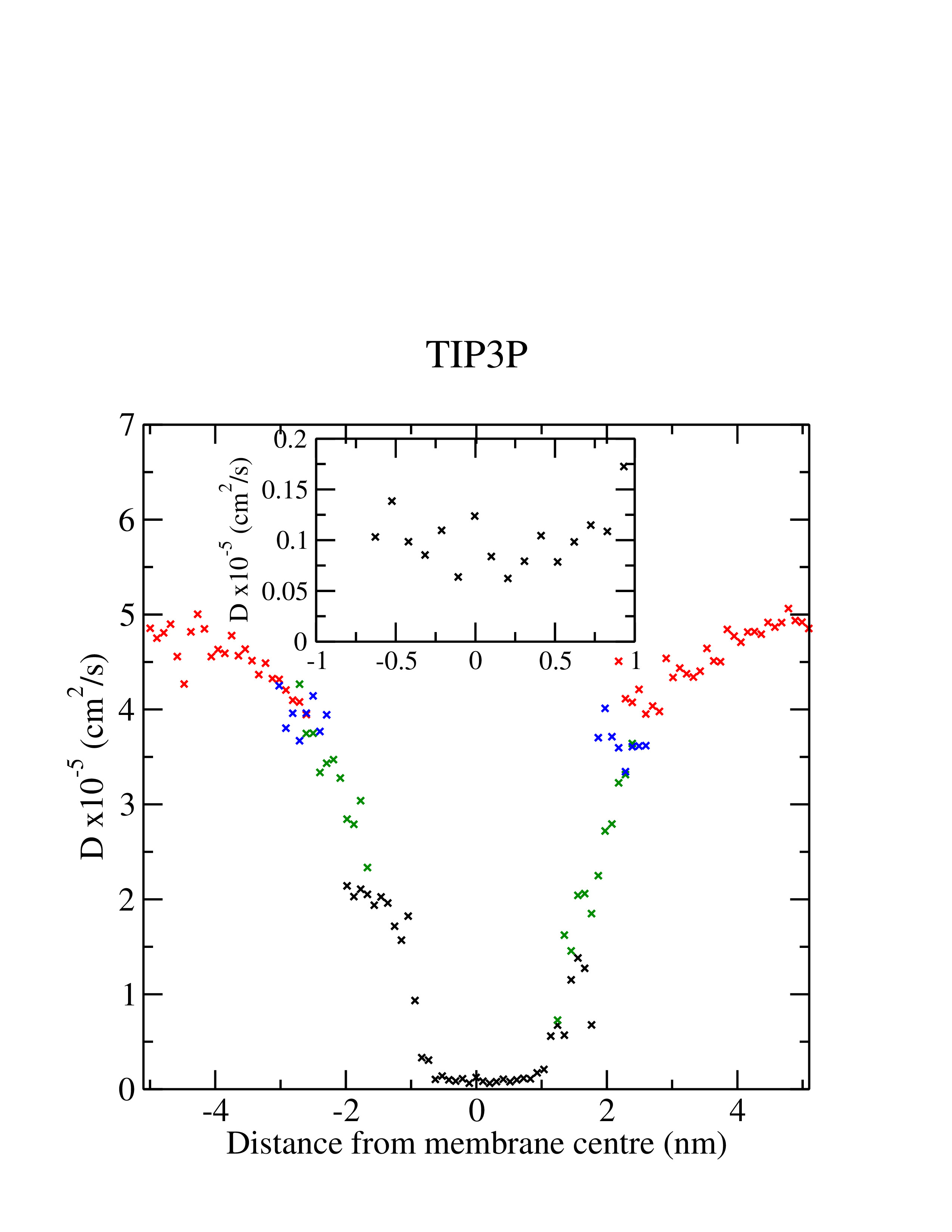}}
\hspace{\fill}
\subfloat{
\includegraphics[clip,width=0.3\textwidth]{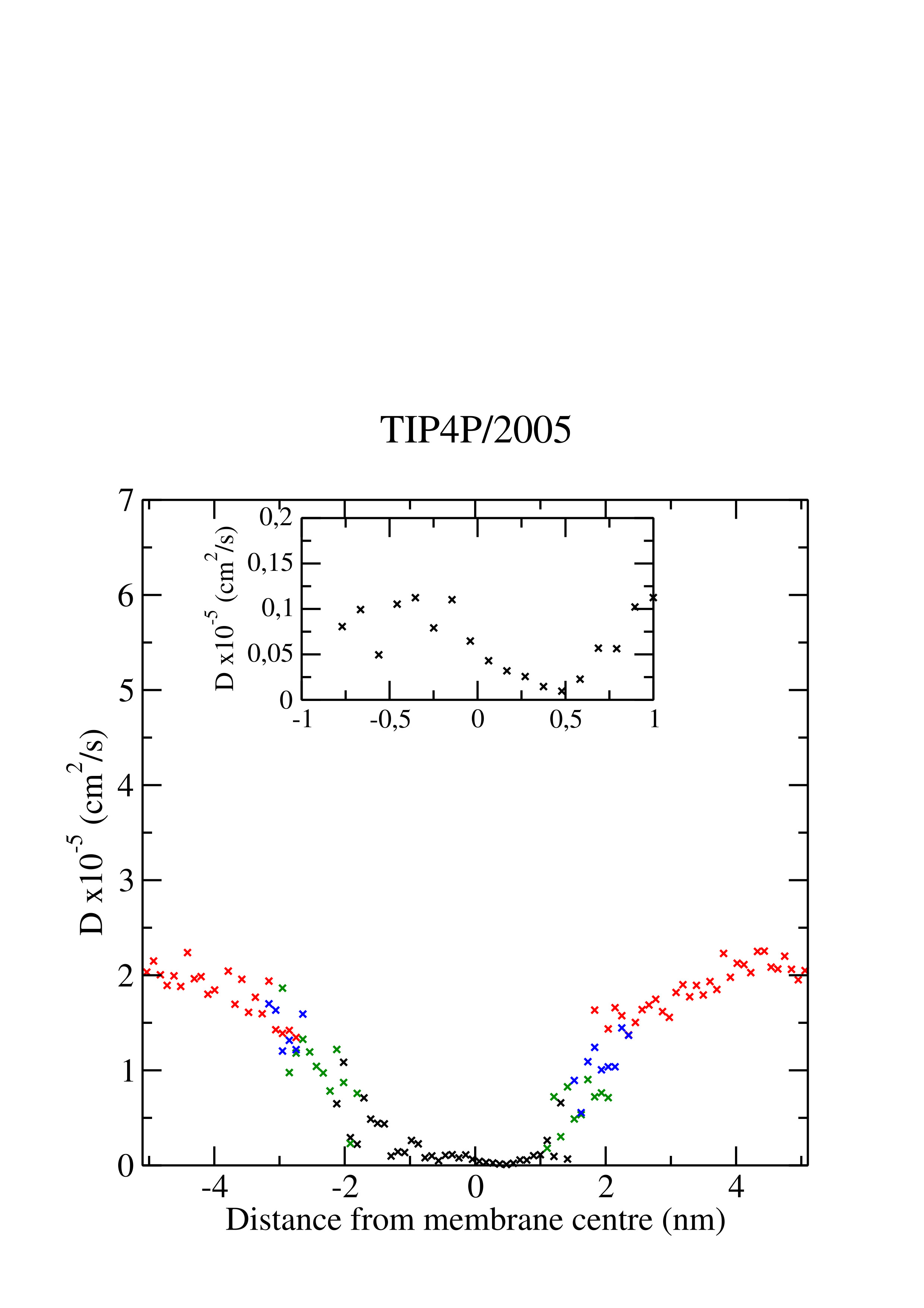}}
\hspace{\fill}
\subfloat{
\includegraphics[clip,width=0.3\textwidth]{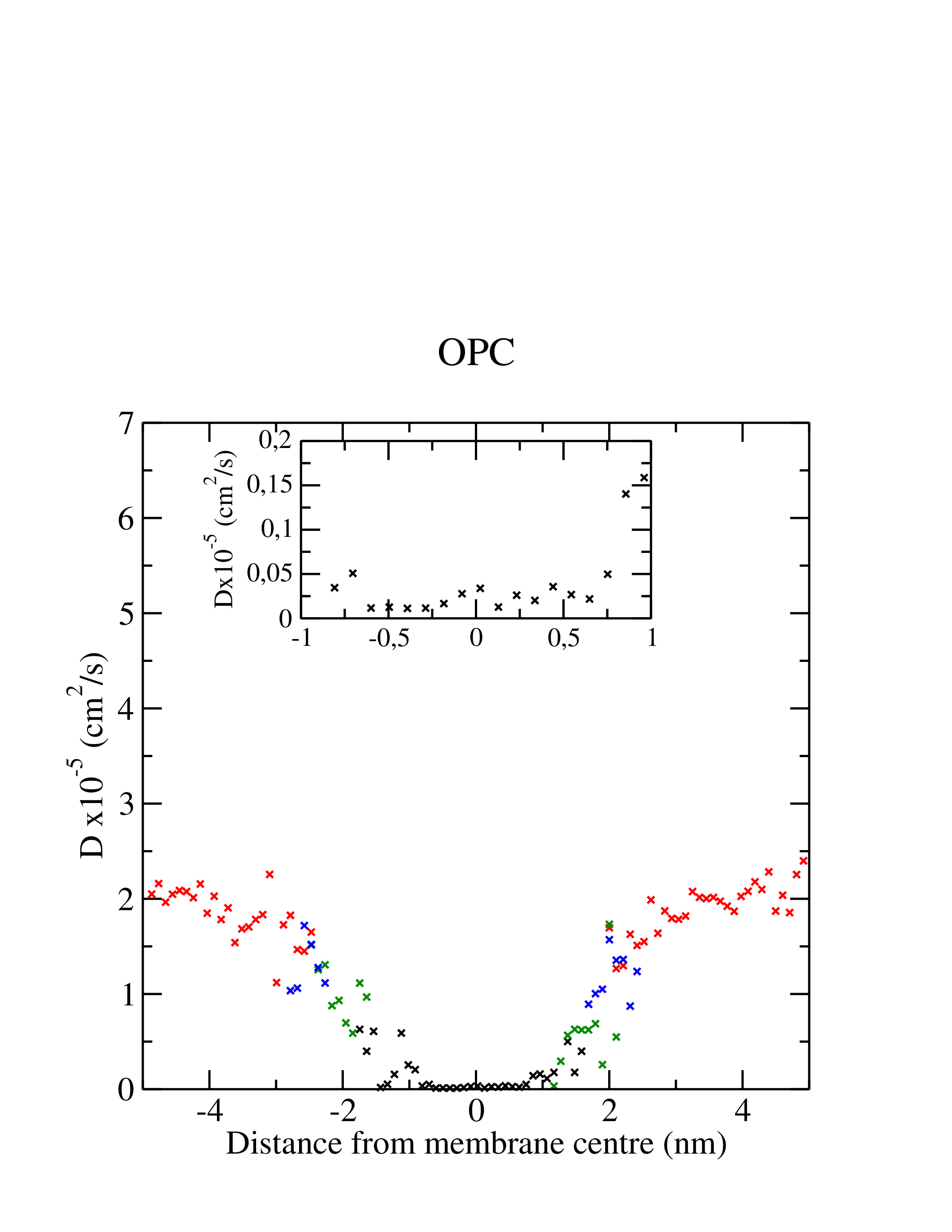}}\\
\caption{Diffusion profiles for TIP3P (left), TIP4P/2005 (centre), and OPC (right) water models. The profiles have been split into four regions: bulk (red crosses), interface (blue crosses), lipid-water contact (green crosses), and AQP region (black crosses).
}
\label{fig:DiffProfiles}
\end{figure*}

In order to visualise the values of $D$ across the four regions, a diffusion profile has been plotted for each water model, see Fig.~\ref{fig:DiffProfiles}.
This shows the diffusion profiles for the three models by considering only the molecules that have crossed the membrane through the protein channel, as these molecules allow us to compute the water diffusion through the channel.

%%%%%%%%%%%%%%%%%%%%%%%%%%%%%%%%%%%%%%%%%%%%%%%%%%%%%%%%%%%%%%%%%%%%%%
%%%%%%%%%%%%%%%%%%%%%%%%%%%%%%%%%%%%%%%%%%%%%%%%%%%%%%%%%%%%%

It can be observed  that $D$ decreases from the bulk region to the AQP region, starting from a value of $D$ lower than $D_{literature}$.
Each region displays a distinct diffusion profile, although there is a slight overlap between regions.
This is due to the criteria used to define the regions, \textit{i.e.} an average of the phosphorus atoms position in $z$. 
As a consequence, it is possible to find a water molecule having a lower value of $D$ than the one expected in that region. 
As shown  in Fig.~\ref{fig:DiffProfiles}, the AQP region is not symmetric and there are higher values of $D_{PORE}$ at negative  distances, $d$, from the membrane centre than at positive ones.
This is because the membrane protein is not symmetric and the diffusion around the protein is asymmetric as a consequence.

\subsection{Diffusion permeability and osmotic permeability}
\label{subsec:permeability}

To compute the water permeability, both the diffusion  (p$_{d}$) and osmotic permeability (p$_{f}$), 
we have calculated the rate at which  molecules  are able to cross the membrane ($q_{0}$ and $k_{0}$) and the diffusion of those molecules through the AQP ($D_{z}$). 
The results have been collected in Table~\ref{tab:permeability}, where we compare numerical and experimental data.
\begin{table*}
\caption{\label{tab:permeability}Number of molecules per second (q$_{0}$, multiplied by 10$^{8}$ (s$^{-1}$) ; diffusion permeability p$_{d}$, by 10$^{-14}$ (cm$^{3}$/s) ; osmotic permeability p$_{f}$, by 10$^{-14}$ (cm$^{3}$/s) ([*] via $p_{f}/p_{d}=11.9$); number of molecules per second for an osmotic process k$_{0}$, by 10$^{9}$ (s$^{-1}$); water diffusion inside the AQP $D_{z}$, by 10$^{-5}$ (cm$^{2}$/s) ([**] via Eq.~\ref{eq:diffu_z}) and $D_{PORE}$, by 10$^{-5}$ (cm$^{2}$/s) }
%\centering
%\begin{threeparttable}
\begin{ruledtabular}
\begin{tabular}{lcccccc}
\hline
Water model & q$_{0}$  & p$_{d}$  & p$_{f}$ ~$^{*}$ & k$_{0}$ ~$^{**}$ & D$_{z}$~$^{**}$ & D$_{PORE}$ \\ 
\hline
TIP3P      & 4.0 (0.3) &  1.20 (0.09) &  14.2 (1.1) & 4.85 & 0.190 & 0.10 (0.01)\\%18\\
TIP4P/2005 & 0.65 (0.2) & 0.194 (0.07) & 2.31 (0.8) & 0.77 & 0.03 & 0.05 (0.02)\\%3 \\
OPC        & 0.5 (0.1) &  0.149 (0.04) & 1.78 (0.4) & 0.59 & 0.02 & 0.02 (0.02) \\
Exp.       &  & & 5.43~\cite{Walz1994} & & & 0.04\\    

\end{tabular}
\end{ruledtabular}
\end{table*}

Diffusion permeability ($p_{d}$) has been computed using Eq.~\ref{eq:osmopermea2}, which is based on the number of water molecules that crossed the membrane through the membrane protein ($q_{0}$) and the volume of a single water molecule ($\nu_{w}=2.989$x$10^{-23}$~cm$^{3}$).
From molecular dynamics simulations $q_{0}$ is a parameter that is easily accessible.
For TIP3P $q_{0}=4.0$x10$^{8}$~s$^{-1}$ whereas for TIP4P/2005 ($q_{0}=0.65$x10$^{8}$~s$^{-1}$) and OPC ($q_{0}=0.5$x10$^{8}$~s$^{-1}$) the number of molecules per second is lower, by almost an order of magnitude.
Therefore, we find 1) on the one hand an overestimation of the TIP3P  diffusion permeability and 2) on the other hand a good performance of both TIP4P/2005 and OPC  diffusion permeability (see Table~\ref{tab:permeability}).

The osmotic permeability $p_{f}$ has been computed from the diffusion permeability using the ratio $p_{f}/p_{d}=N+1=11.9$ proposed by Zhu \textit{et al.}~\cite{Zhu2004} for AQP1 and the values are reported in Table~\ref{tab:permeability}.
The experimental data for $p_{f}$ is $p_{f}=5.43$x $10^{-14}$ cm$^{3}$/s~\cite{Walz1994}, which is in very good agreement with the TIP4P/2005 (2.31x$10^{-14}$cm$^{3}$/s) and OPC ($1.78$x$10^{-14}$cm$^{3}$/s) values. The TIP3P value, 
14.2x$10^{-14}$cm$^{3}$/s,  is dramatically higher than the experimental value.

Having calculated the osmotic permeability, the rate of water crossing the AQP ($q_{0}$) can be calculated by applying Eq. 2 with a value of z = 0.28 nm ~\cite{Horner2015}.
From the same Eq.~\ref{eq:diffu_z}, we calculated $D_{z}$ for the different water models, obtaining 19.0x10$^{-7}$~cm$^{2}$/s, 3.03x10$^{-7}$~cm$^{2}$/s, and 2.33x10$^{-7}$~cm$^{2}$/s for TIP3P, TIP4P/2005 and OPC, respectively.
TIP4P/2005 and OPC compare well with the experimental data, 4.0~x10$^{-7}$~cm$^{2}$/s.
However, TIP3P shows a value higher  with respect to the experimental one  than the other water models.

It is worth nothing that $D_{z}$ could be compared to $D_{PORE}$ using  Eq.~\ref{eq:diffu_constant} as $dim=1$.
This fact allows us to test the consistency of our diffusion results, since $D_{PORE}$ is calculated via Eq.~\ref{eq:diffu_constant} and $D_{z}$ is computed via an independent method (via Eq.~\ref{eq:diffu_z}) based on the number of water molecules that crossed the membrane through AQP ($q_{0}$) and its ratio with respect to the water permeability.
As it can be observed, the results for water diffusion inside the AQP are in excellent agreement independently of the method used to compute it, either using Eq.~\ref{eq:diffu_constant} on the AQP region or applying Eq.~\ref{eq:diffu_z}.

In summary, our results provide compelling evidence that the TIP3P water model shows a higher diffusion rate than the other water models studied. Inside the channel, the TIP4P/2005 model produces results in excellent agreement with experiment.
Focusing on the diffusion through the AQP protein using two independent methods, we demonstrate that there is an overestimation of the diffusion, resulting in   overestimated values of water permeability for TIP3P.
However, for TIP4P/2005 and OPC, which have bulk diffusion values close to the experimental one, their calculated values of water permeability are comparable to the experimental value. This implies that TIP4P/2005 and OPC are capable of producing reliable estimates of water diffusion  through the AQP1 channel.

\section{Conclusion}

In this work, water diffusion and  permeability have been studied using molecular dynamics for three systems,  consisting of AQP1 embedded in a POPC lipid membrane and three water models, TIP3P, TIP4P/2005 and OPC.
TIP3P is the standard water model used for CHARMM36m, whereas TIP4P/2005 and OPC are water 
models not frequently used in biomolecular simulations but  known to better reproduce bulk water properties~\cite{vega11,Izadi2014}.

Recently a number of studies have  addressed the relevance of water models in e.g. protein aggregation and host-guest recognition systems  \cite{cinaroglu2021,emperador2021effect}.
Our main purpose was to compare among the behaviour of different water models and shed light on  AQP1 water permeability. 
This enabled us to obtain reliable results for the molecular mechanism of water diffusion as well as both diffusion and permeability values,  near the membrane and through the AQP1.
%This would  give reliable results of the molecular mechanism, water diffusion, both near the membrane and through the AQP1, and water permeability. 
%distinguishing between diffusion permeability and osmotic permeability. 
First of all, the well-known molecular mechanism of water permeation through AQP1 has been tested for the three models, giving similar and excellent results.
All of them show a specific orientation of water molecules inside the AQP1 as demonstrated in Ref.~\cite{Tajkhorshid2002}.
Hence, using any of these water models, the water-protein interaction is reliable and AQP1 keeps its high-selectivity permeation of water across the membrane.

The obvious difference among water models can be detected in the transport properties. 
Diffusion inside the protein has been computed using two methods, one based on the mean square displacement considering the geometry of the channel and another one from the water permeability, giving consistent results for the three models.
TIP3P has a diffusion constant twice that of the experimental value, whereas TIP4P/2005 and OPC produce values remarkably close to  experiment.
%while TIP4P/2005 and OPC reproduce remarkable well values of $D$ compared to the experimental value.
The overestimation of TIP3P is consistent in the whole process of water permeation across the membrane.
The most relevant region is inside the membrane protein, \textit{ i.e.}, the AQP region, where the TIP3P diffusion is higher than the experiments. On the other hand, OPC presents a lower diffusion rate. When comparing to experiments, we conclude that TIP4P/2005 is the model that best resembles water diffusivity inside the channel.
It is worth noting that the ratios of TIP3P/TIP4P/2005 and TIP3P/OPC increase when moving from bulk to the interfacial region to the AQP region.  Therefore, the difference between the models seems to be more pronounced within the nanopore environment.
%in the pore environment. 
%Diffusion inside the protein has been computed using two methods, one based on the mean square displacement considering the geometry of the channel and another one from the water permeability, giving consistent results for the three models.

Focusing on water permeability, we have computed the number of water molecules  able to cross the membrane through the AQP1, $q_{0}$. By means of  this parameter we calculated the diffusion permeability, which allowed us to obtain the value of osmotic permeability.
As in the diffusion case, TIP3P gives a diffusion permeability between 6 and 8 times higher than TIP4P/2005 and OPC, respectively.

The overestimation of TIP3P is observed also in the calculation of the  osmotic permeability (being almost 3 times higher than the experimental counterpart) and in the water diffusion inside the AQP1 (being 5 times higher than the experimental counterpart). However the calculated values obtained with the OPC and TIP4P/2005 water models are in much better  agreement with the experiments.
%In conclusion, the results of osmotic permeability computed for AQP1 in contact with TIP3P might are higher than experiments dues to the demonstrated overestimation of water diffusion for TIP3P.
It is difficult to compute the osmotic permeability in experiments and there are several values of $p_{f}$. 
However, in this work reliable values of water diffusion and water permeability are reported, which demonstrate the clear differences between the behaviour of TIP3P and the results of TIP4P/2005 and OPC.
We believe that this observation can be relevant not only for the AQP channel but also for numerical studies of transport properties of water and ions within nanopores and ion channels.
%ions in the presence of ionic channels. 
%water and/or ion diffusion in nanopores
%\textcolor{red}{Rewrite this idea: This is relevant not only for AQP but for other simulation studies of water and/or ion diffusion in nanopores.}

Simulating polarizable models require a considerable increase in computational effort, and to date their application has generally been limited to somewhat simpler membrane~\cite{Chen2021} and channel systems~\cite{Klesse2020,Ngo2021}. We therefore have decided to first understand the basic features of water transport through AQP1, using only  non-polarizable models. However, in the future it will be important  to consider systematically polarizable models for water, proteins and lipids \cite{ren2003polarizable,Ngo2021}.

This study has focused on the dynamical properties of the water model in AQP simulations. However, it should be noted that the protein and lipid parameters have been optimized for use with e.g. TIP3P. The question therefore remains whether in the future one should re-optimize these parameters for improved water models or switch to polarizable models. The latter are being extended to lipids as well as to water and proteins and are now feasible, if computationally expensive, for membrane channel protein systems.\cite{Jing2021}
An interim compromise may be the use of charge scaling (via the electronic continuum correction approach) to mimic electronic polarization, which has been used with some success for biological membranes.\cite{Duboue-Dijon2020}

\section*{Supplementary material}
See  the  supplementary  material is available free of charge at 
\begin{itemize}
    \item Dipole moment for TIP4P/2005. (Figure S1)
\end{itemize}

\section*{DATA AVAILABILITY}
The data that support the findings of this study are available from the corresponding author upon reasonable request.

\section*{acknowledgements}

The work has been performed under the Project HPC-EUROPA3 (INFRAIA-2016-1-730897), with the support of the EC Research Innovation Action under the H2020 Programme; in particular, the author gratefully acknowledges the support of M.S.P. Sansom and the computer resources and technical support provided by ARCHER. M.A.G. thanks for the support by Ayuda Juan de la Cierva-Incorporaci\'on (IJCI-2016-27497) from Ministerio de Ciencia, innovaci\'on y universidades (Spain). 
M.S.P. Sansom  acknowledges the following fundings: BBSRC BB/N000145/1
EPSRC EP/R004722/1 and EP/V010948/1, and C. Valeriani acknowledges fundings from MINECO PID2019-105343GB-I00.

%\bibliography{mac,MEMBRANE,Mios,AQP1}
%aipnum4-2.bst 2019-01-14 (MD) hand-edited version of apsrev4-1.bst

\end{document}